\documentclass{article}
\usepackage{mrs2005,epsfig}
\begin{document} 
\title{SUPERNOVA OUTFLOWS IN GALAXY FORMATION}

\author{C. SCANNAPIECO$^{1,2}$, P.B. TISSERA$^{1,2}$, S.D.M. WHITE$^3$, V. SPRINGEL$^3$}
\affil{$^1$ Instituto de Astronom\'{\i}a y F\'{\i}sica del Espacio, Casilla de Correos 67,
Suc. 28, 1428, Buenos Aires, Argentina.\\
$^2$  Consejo Nacional de Investigaciones Cient\'{\i}ficas
y T\'ecnicas, CONICET, Argentina.\\ 
$^3$  Max-Planck Institute for Astrophysics, Karl-Schwarzchild Str. 1, D85748, Garching, Germany.}

\begin{abstract} 

We investigate the generation of galactic outflows by supernova
feedback in the context of SPH cosmological simulations. We use a
modified version of the code {\small GADGET-2} which includes chemical
enrichment and energy feedback by Supernova.  We find that energy
feedback plays a fundamental role in the evolution of galaxies,
heating up the cold material in the centre of the haloes and
triggering outflows which efficiently transport gas from the centre to
the outskirts of galaxies.  The impact of feedback is found to depend
on the virial mass of the system with smaller systems, such as dwarf
galaxies, being more strongly affected.  The outflows help to establish
a self-regulated star formation process, and to transport a
significant amount of metals into the haloes and even out of the
systems.  According to our results, energy feedback by supernovae
could be the mechanism responsible for the chemical enrichment of the
intergalactic medium.

\end{abstract} 
 
\section{Introduction}

The important impact of supernova (SN) explosions on the evolution of
galaxies has been discussed by numerous numerical works during the
last years (e.g. Katz \& Gunn 1991; Cen \& Ostriker 1992, 1999; Yepes
et al. 1997).  On one hand, SNe constitute the main source of heavy
elements in the Universe, and the presence of metals in diffuse gas
can accelerate the gas condensation in potential wells, since the
cooling rate of baryons depends sensitively on metallicity (Sutherland
\& Dopita 1993). On the other hand, the injection of energy into the
interstellar medium by SN explosions provides an efficient mechanism
to heat up material in the centre of the haloes and to transport it to
the outskirts of galaxies.  These {\it outflows} are thought to play a
crucial role in the subsequent evolution of galactic systems, helping
to establish a self-regulated star formation process and triggering a
redistribution of mass and metals.

Observationally, it is found that the intergalactic medium is
contaminated with metals. Since metals are produced in stellar
interiors and ejected into the star forming regions deep inside
galactic potential wells, an efficient mechanism for transporting
metals to the low-density intergalactic medium is needed in order to
explain the observed contamination. Strong SN outflows constitute a
natural explanation for the presence of metals in the intergalactic
medium.

We use numerical Smoothed Particle Hydrodynamics (SPH; Gingold \&
Monaghan 1977; Lucy 1997) simulations which take into account chemical
enrichment and energy feedback by SN in order to study the impact of
SN explosions on the evolution of galaxies and the chemical enrichment
patterns of the stellar populations, as well as the gas in the
interstellar and the intergalactic media.  Numerical codes including
chemical and energy feedback have already become an important tool for
studying the effects of SN feedback in the cosmological context.
However, SPH codes have encountered severe complications in
realistically accounting for SN explosions, with the energy feedback
being particularly difficult to describe.  Previous works have shown
that a simple injection of thermal energy into the cold gas has only a
small effect on the hydrodynamics, as a consequence of the short
cooling times of the dense cold gas, which thermalizes the feedback
energy very quickly (e.g.  Kay et al. 2002; Marri \& White 2003;
Springel \& Hernquist 2003).  We have developed a new model which is
aimed at improving the description of SN feedback, allowing us to
overcome some failures of previous approaches without introducing
any scale-dependent parameter (Scannapieco et
al. 2005a,b). In this work, we analyse the effects of SN explosions on
the evolution of isolated disc-type systems, focusing on the star
formation activity and the impact of the outflows on the dynamical and
chemical evolution of galaxies.

This article is organized as follows. In Section~\ref{model}, we
summarise the main characteristics of our model for chemical
enrichment and energy feedback by supernova.  In
Section~\ref{results}, we analyse the properties of galactic objects
of different virial mass, in relation to the star formation activity
and metal distribution. Finally, in Section~\ref{conclusions} we give
our conclusions.

\section{Numerical Models}\label{model}

We use a novel model for chemical enrichment implemented in the code
{\small GADGET-2} (Springel \& Hernquist 2003, Scannapieco et
al. 2005a,b) in order to study the impact of supernova feedback on the
formation of galaxies.  The code includes metal-dependent cooling,
star formation, chemical enrichment and energy feedback by Type II and
Type Ia supernovae. The model is tied to an explicit multiphase
treatment (Marri \& White 2003) for the gas components and includes a
treatment of a cosmological UV background.

Here we briefly describe our model for chemical enrichment and energy
feedback.  The interested reader is refered to Scannapieco et
al. (2005a,b) for details. At each integration time-step of the code,
we calculate the number of exploding Type II and Type Ia supernovae,
accounting for the associated chemical and energy production.  We
assume that SNII explode within a time-step of integration of the
code, while SNIa life-times are selected randomly from a given range
(Scannapieco et al. 2005a). Each SN injects $10^{51}$ ergs of energy
into the surrounding gas.  Our model for SN feedback is based on an
explicit segregation of the gas surrounding an exploding star particle
into a cold-dense phase and a diffuse phase for the purpose of
releasing its energy and chemical production.  These phases are
loosely refered to as {\it cold} and {\it hot}, and are defined in the
following manner: the cold phase is formed by the gas with $T < 2\,
T_*$ and $\rho > 0.1 \rho_*$ ($T_* = 4 \times 10^4$ K, $\rho_* = 7
\times 10^{-26}$ g cm$^{-3}$), while the rest of the gas is considered
to be part of the hot phase. Note that these phases are determined
only for exploding star particles at the time of metal and energy
distribution.  In our model, we assume that the feedback energy is
dumped in given proportions into the cold and the hot phases of the
exploding stars: a fraction $\epsilon_c$ is dumped into the cold phase
while the remaining $\epsilon_h = 1 -\epsilon_c$ goes into the hot
one.  For the hot phase, the injection of feedback energy is
simultaneous with the explosion. In contrast, for each particle in the
cold phase, we define a reservoir in which the energy received by
successive explosions is accumulated. When this energy is sufficient
for the particle to reach the entropy of its hot neighbours, the
particle is {\it promoted} to the hot phase, and the reservoir energy
is dumped into its internal energy. This promotion scheme drives a
transformation of gas from the cold-dense phase to the hot-diffuse one
(Scannapieco et al. 2005b).  Note that the scheme for energy feedback
does not introduce any ad-hoc parameter, and there is only one free
parameter to assume: $\epsilon_c$ (or equivalently $\epsilon_h$).
Metals are also distributed into the hot and cold phases of the
exploding stellar particles accordingly to the corresponding fractions
($\epsilon_h$ or $\epsilon_c$).  However, the chemical release occurs
always simultaneously with SN explosions.

The simulations also include a multiphase treatment for the gas
component similar to the model of Marri \& White (2003), which is
based on an explicit decoupling of the gas with different
hydrodynamical properties. Within the scheme, we explicitly prevent
gas particles of being neighbours if their entropies differ by more
than a given factor. The model is found to be insensitive to the value
used for the decoupling within physically motivated limits.  In our
model, a better defined hot phase is allowed to form, even within cold
regions (Scannapieco et al. 2005b).

\section{Results}\label{results}

In order to investigate the performance of our SN feedback model, we
simulate the evolution of isolated disc-type galaxies. The initial
conditions are generated by radially perturbing a spherical
distribution of superposed dark matter and gaseous particles in order
to give rise to a density profile of the form $\rho(r)\sim r^{-1}$.
The sphere is initially in solid body rotation with an angular
momentum characterised by a spin parameter of $\lambda \approx 0.1$.
We have simulated systems of both $10^{12}$ and $10^{10}$ M$_\odot$
$h^{-1}$ virial mass ($h = 0.7$), 10 per cent of which is in the form
of baryons, and with a $\sim 9000$ particle resolution for both the
gaseous and the dark matter components.  For the $10^{12}$ M$_\odot$
$h^{-1}$ system, this yields a mass resolution of about $\sim 10^8$
M$_\odot h^{-1}$ and $10^7\ $ M$_\odot h^{-1}$ for the dark matter and
gas particles, respectively.  We adopt maximum gravitational
softenings of $1.50$, $0.75$ and $1.13$ kpc $h^{-1}$ for dark matter,
gas and stars, respectively.  For the $10^{10}$ M$_\odot $ $h^{-1}$
mass system, the softenings were correspondingly rescaled.

\begin{table}
\center
\caption{Main characteristics of the test simulations: virial mass,
input feedback parameters $\epsilon_c$ and $\epsilon_h$ (for the
feedback tests $\epsilon_h=1-\epsilon_c$), and mass of gas with
positive total energy, normalized to total baryonic mass (unbound gas
fraction, $f_{\rm unbound}$).}
\vspace{0.25cm}
\begin{tabular}{cccc }
M$_{\rm vir}$ [$ {\rm M}_\odot\ h^{-1} $] &  $\epsilon_c$ &$\epsilon_h$ &$f_{\rm unbound}$\\\hline
&&&\\
  $10^{12}$  &     0          &  0   &    0.02   \\
  $10^{12}$  &     0.1        &  0.9 &    0.34   \\
  $10^{12}$  &     0.5        &  0.5 &    0.56   \\
  $10^{12}$  &     0.9        &  0.1 &    0.53   \\
&&&\\
  $10^{10}$  &     0          &  0   &    0.00    \\
  $10^{10}$  &     0.1        &  0.9 &    0.98    \\
  $10^{10}$  &     0.5        &  0.5 &    0.62    \\
  $10^{10}$  &     0.9        &  0.1 &    0.87    \\
\end{tabular}
\label{table_unbound}
\end{table}

For all tests, we have used the same star formation and SN parameters:
a star formation efficiency of $c=0.1$, a SNIa rate of 0.001 relative
to SNII and a life-time interval for SNIa of [0.1,1] Gyr. In order to
test the dependence of the model on the feedback parameter
$\epsilon_c$, which sets the fraction of energy and metals distributed
into the cold phase, we have performed simulations with
$\epsilon_c=0.1, 0.5$ and $0.9$. For comparison, we have also run
simulations without including the energy feedback treatment.  In
Table~\ref{table_unbound}, we summarise the main feedback parameters
of the simulations.

\begin{figure}  
\vspace*{1.25cm}  
\begin{center}
\epsfig{figure=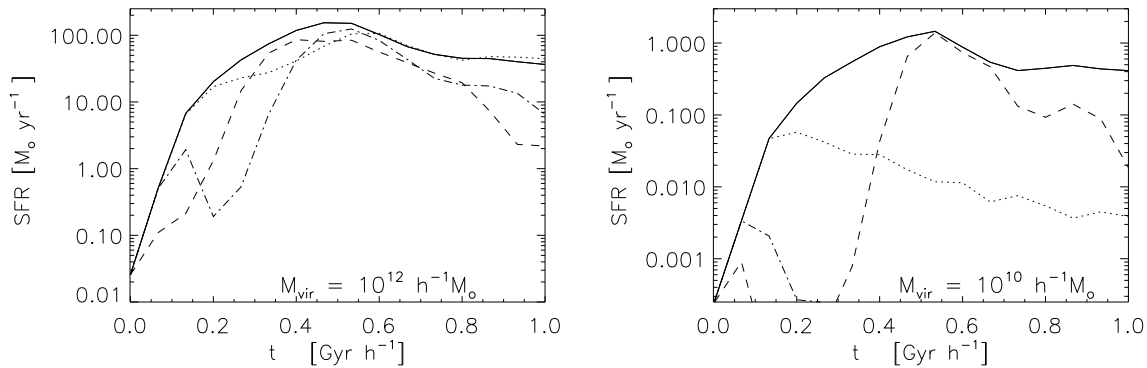,width=15.5cm}  
\end{center}
\vspace*{0.25cm}  
\caption{Star formation rate for our tests of $10^{12}$ (left panel)
and $10^{10}$ (right panel) M$_\odot$ $h^{-1}$ virial mass galaxies,
for runs with no feedback (solid lines) and with feedback and
different input parameters: $\epsilon_c=0.1$ (dotted lines),
$\epsilon_c=0.5$ (dashed lines) and $\epsilon_c=0.9$ (dotted-dashed
lines). }
\label{sfr}
\end{figure}

In the left panel of Fig.~\ref{sfr}, we show the star formation rates
for our $10^{12}$ ${\rm M}_\odot$ $h^{-1}$ virial mass systems. From
this figure we can see that the inclusion of feedback always leads to
a decrease in the star formation activity. This is a consequence of
the injection of energy by SNe which heats up the gas surrounding the
explosions. As a result, the density of the cold gas from which stars
are formed decreases, reducing in turn the star formation activity.
We can also see that as we inject a larger fraction of the energy into
the cold phase (i.e. as the $\epsilon_c$ value increases), the
decrease in the star formation rate is stronger. In the extreme case
of $\epsilon_c=0.9$, the star formation rate is significantly reduced
after the formation of the first stars. Later on, the gas is able to
cool down again, triggering a starburst.  We find that the final
stellar mass fraction of the test run with $\epsilon_c=0.1$ is $20$
per cent lower than its counterpart in the test without feedback,
while tests with $\epsilon_c=0.5$ and $0.9$ lead to a reduction of
$\sim 50$ per cent with respect to the no feedback test.

In the case of the $10^{10}$ ${\rm M}_\odot\ h^{-1}$ system, we find
that the effects of feedback are more extreme, as can be seen from the
right panel in Fig.~\ref{sfr}.  In this case, there is a stronger
dependence of the results on the input feedback parameter. For
$\epsilon_c=0.1$ (i.e.  $10$ per cent of the feedback energy is
injected into the cold phase), most of the hot gas is accelerated
outwards before it can collapse.  The gas that is able to cool gives
rise to the remaining level of star formation, which leads to an
almost negligible final stellar mass fraction. Note that in this case,
the cold gas receives only $10$ per cent of the feedback energy, so it
is unlikely that it can receive enough energy to be heated up.  In the
test with $\epsilon_c=\epsilon_h=0.5$, the energy injected after the
first stars helps to reduce the star formation rate but the system is
able to cool down and collapse later.  In this case, the interplay
between radiative cooling and heating by SN leads to a series of
starbursts, setting a self-regulated mechanism for the star formation.
This bursty behaviour is consistent with observational results of
dwarf galaxies which find that the star formation is delayed in such
low mass systems and recent starbursts can develop (Kauffmann et
al. 2003).  Finally, if the bulk of the SN energy is injected into the
cold phase ($\epsilon_c=0.9$), most of the gas is heated up and swept
away from the system very quickly, leading to a negligible final
fraction of stars.  It is interesting to note that if we inject most
of the energy either into the cold or the hot phase, the final stellar
mass formed is very small in both cases.

So far we have analysed how the star formation rates can be
self-regulated depending on the virial mass of the systems as a result
of the injection of energy triggered by SN explosions.  Here, we
discuss the strength of the outflows generated by the feedback.  For
this purpose we calculate the unbound gas fraction, defined as the
mass of gas with positive total binding energy (kinetic plus
gravitational) normalized to the total baryonic mass, as an estimator
of the feedback strength.  We find that the fraction of unbound gas
varies with the details of the SN energy injection and the mass of the
system.  In Table~\ref{table_unbound}, we show the fraction of unbound
gas, normalized to total baryonic mass, from our tests of both virial
mass and for different choices of the feedback parameter, after $0.7$
Gyr $h^{-1}$ of evolution. From the results of the $10^{12}$ M$_\odot$
$h^{-1}$ virial mass system we can see that the unbound gas fraction
increases with the value of $\epsilon_c$, so the fraction of unbound
material increases when a larger fraction of the energy is dumped into
the cold phase.  In these tests, regardless of the $\epsilon_c$ value,
we find that $\sim 30$ per cent of the unbound gas has $|v_z| > 500$
km s$^{-1}$ after the starburst, where $|v_z|$ is the absolute value
of the $z$-velocity.  Note that if energy feedback is not included,
there is no efficient mechanism to transport material out of the
system, and the fraction of unbound gas is negligible.  In the case of
the $10^{10}$ M$_\odot$ $h^{-1}$ virial mass system, we find that the
test with $\epsilon_c=0.5$ has an unbound gas fraction of $60$ per
cent, while in the other feedback tests $\sim 90$ per cent of the gas
content is expelled from the system, preventing the gas
collapse. Hence, in our model the strength of the outflows depends on
the virial mass as well as on the input feedback parameter.

\begin{figure}  
\vspace*{1.25cm}  
\begin{center}
\epsfig{figure=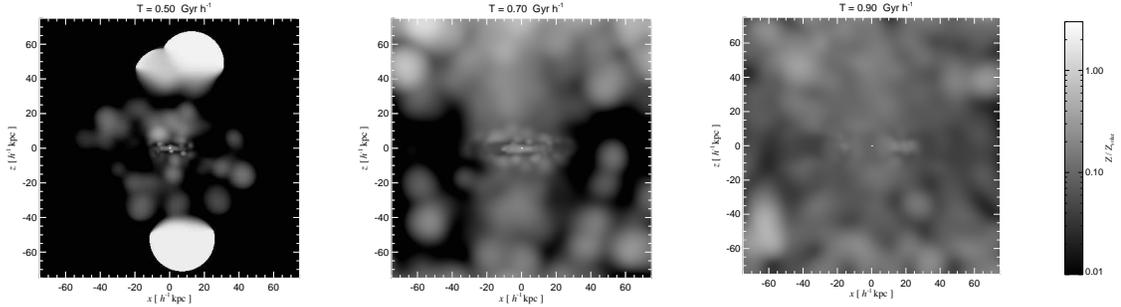,width=14cm}  
\end{center}
\vspace*{0.25cm}  
\caption{Edge-on metallicity maps for our test of a $10^{12}$
M$_\odot$ $h^{-1}$ virial mass galaxy run with energy feedback and
$\epsilon_c=0.5$, at different times during the evolution.  The
metallicity scale is also shown.}
\label{maps}
\end{figure}

The outflows constitute an effective mechanism to transport metals to
the outskirts of galaxies.  Note that the outflows are generated in
the inner parts of the systems where the gas is highly
contaminated. Hence, if outflows are able to transport a significant
fraction of the gas outwards, they can contaminate the haloes.  In
Fig.~\ref{maps}, we show metallicity maps for our test of the $10^{12}$
${\rm M}_\odot\ h^{-1}$ system run with $\epsilon_c=0.5$, for three
different times. It is clear from this figure that metals are
effectively transported outwards in our model, enriching the outer
parts of the systems.  Note that the outflows are mostly perpendicular
to the disc plane.

\begin{figure}  
\vspace*{1.25cm}  
\begin{center}
\epsfig{figure=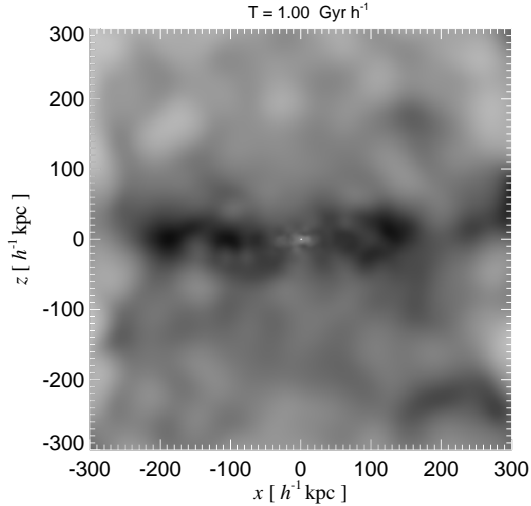,width=7cm}  
\end{center}
\vspace*{0.25cm}  
\caption{Extended metallicity map for our test of $10^{12}$ M$_\odot$
$h^{-1}$ virial mass galaxy run with energy feedback and
$\epsilon_c=0.5$, after $1$ Gyr $h^{-1}$ of evolution.  The
metallicity scale is shown in Fig.~\ref{maps}.}
\label{map-300kpc}
\end{figure}

Based on our results, we argue that energy feedback could be
responsible for the existence of metals in the intergalactic
medium. As an example, we show in Fig.~\ref{map-300kpc} a metallicity
map up to $300$  kpc $h^{-1}$, for our test of $10^{12}$ ${\rm
M}_\odot\ h^{-1}$ mass and $\epsilon_c=0.5$, after $1$ Gyr $h^{-1}$ of
evolution.  From this figure, we can see that enriched gas has been
effectively transported outwards and has even escaped from the
potential well of the system.  From these initial conditions of
isolated galaxies we cannot reliably constrain the spatial extent of
the contamination. We plan to study cosmological simulations for this
purpose (Scannapieco et al. 2005c).

\section{Conclusions}\label{conclusions}

We have analysed the impact of supernova explosions on galaxies using
SPH cosmological simulations which include chemical enrichment and
energy feedback. We have run simulations of isolated galaxies from
idealized initial conditions of both $10^{10}$ and $10^{12}$ M$_\odot$
$h^{-1}$ virial mass, and studied their star formation histories and
metal distributions.

Our model for SN feedback is able to heat up material in the centre of
the systems and to produce substantial outflows of gas. For a given
feedback input parameter (which sets the fraction of energy
dumped into the cold gas), 
the strength of the outflows is found to depend on the virial 
mass.

The outflows drive a redistribution of the gas, lifting cold material
from the centre of the haloes, heating it up and transporting it to
the outskirts of galaxies.  As a result, a significant fraction of
the gas is swept away from the centre of the systems, reducing the
cold gas density and consequently the star formation rate.  The
outflows therefore help to establish a self-regulated star formation
process.

According to our results, the effects of SN feedback on the formation
of dwarf galaxies should be crucial.  In our tests of $10^{10}$
M$_\odot$ $h^{-1}$ mass systems, we find that SN feedback is able to
either stop the star formation activity, to reduce it and give rise to
a more or less continuous (and low) level, or to trigger a series of
starbursts, depending on the input feedback parameter.

The outflows triggered by SNe also leave imprints on the chemical
enrichment patterns. Since they are generated in the centre of the
haloes where the material is highly contaminated, the outflows
transport an important fraction of enriched material outwards. This
leads to a chemical contamination of the haloes and could even explain
the existence of metals in the intergalactic medium. In order to study
this process in detail and to predict the large-scale distribution of
metals, it will be necessary to consider fully self-consistent
cosmological simulations.

\acknowledgements{ This work was partially supported by the European
Union's ALFA-II programme, through LENAC, the Latin American European
Network for Astrophysics and Cosmology. Simulations were run on Ingeld
PC-cluster funded by Fundaci\'on Antorchas. We acknowledge support
from Consejo Nacional de Investigaciones Cient\'{\i}ficas y
T\'ecnicas, Agencia de Promoci\'on de Ciencia y Tecnolog\'{\i}a,
Fundaci\'on Antorchas and Secretar\'{\i}a de Ciencia y T\'ecnica de la
Universidad Nacional de C\'ordoba. C. S. thanks the LOC of the meeting for their
financial support.}

\vfill 
\end{document}